\documentclass[epj,final]{svjour}

\usepackage[dvips]{graphicx}
\usepackage{times}

\title{Granular Flows in a Rotating Drum: the Scaling Law
between Velocity and Thickness of the Flow}
\author{Gwena\"elle F\'elix\inst{1} \and V\'eronique Falk\inst{1} 
     \and Umberto D'Ortona\inst{2}}
\institute{%
      Laboratoire des Sciences du G\'enie Chimique, 
      UPR 6811, CNRS, ENSIC, INPL, 
      1 rue Grandville, BP451, 54000 Nancy, France \and
      Laboratoire de Mod\'elisation et Simulation Num\'erique en M\'ecanique et G\'enie
      des Proc\'ed\'es, 38 rue Fr\'ed\'eric Joliot-Curie, 13451 Marseille Cedex 20, France, \email{umberto@l3m.univ-mrs.fr}}

\date{\today}

\abstract{
The flow of dry granular material in a half-filled rotating drum is studied. 
The thickness of the flowing zone is measured for several rotation speeds, drum
sizes and beads sizes (size ratio between drum and beads ranging from 47 
to 7400). Varying the rotation speed, a scaling law linking
mean velocity vs thickness of the flow, $v\sim h^m$, is deduced for each couple
(beads, drum). The 
obtained exponent $m$ is not always equal to 1, value previously reported in 
a drum, but varies with the geometry of the system. For small size ratios, 
exponents higher
than 1 are obtained due to a saturation of the flowing zone thickness.
The exponent of the power law decreases with the size ratio, leading
to exponents lower than 1 for high size ratios. These exponents imply that
the velocity gradient of a dry granular flow in a rotating drum 
is not constant. More fundamentally, these results show that the flow of a 
granular material in a rotating drum is very sensible to the geometry, and that 
the deduction of the ``rheology'' of a granular medium flowing in such a 
geometry is not obvious.
}
\PACS{
{45.50.-j}{Dynamics and kinematics of a particle and a system of particles}\and
{45.70.-n}{Granular systems}
}

\begin{document}
\titlerunning{Granular Flows in a Rotating Drum ...}
\authorrunning{G. F\'elix, V. Falk \& U. D'Ortona}
\maketitle
%
Research in granular material has received a renewed interest these last 
years \cite{OttinoKhakhar01}. Nevertheless, mechanisms governing flows of particles
are not yet completely understood. For example, two 
scaling laws ``mean velocity $v$ vs flowing zone thickness $h$'' have been proposed.
For flows down a rough inclined plane,
Pouliquen \cite{Pouliquen99} and Azanza et al. \cite{Azanza99} have
obtained $v\sim h^{3/2}$ while
Rajchenbach \cite{Rajchenbach00}, Bonamy et al. \cite{BonamyDaviaud02} and
GDR MiDi \cite{GDRMiDi}
report a linear and constant (with the rotation speed) velocity gradient in a 
rotating drum, inducing a scaling law $v\sim h$. 
Recently, Ancey \cite{AnceyPRE02} found, for the flow down an incline, an 
exponent in the scaling law $v\sim h^m$
that is not constant, but presents 2 different values, characteristic of 2 distinct
regimes: for high slopes, $m$ is around 1 to 2 (values that are compatible with 
$m=3/2$, but also
with $m=1$), but for gentle slopes $m$$\simeq$0, which gives the surprising result
that the mean velocity of the flow is constant (independent of the thickness
of the flowing layer).
Except the curvature of the flowing layer, the main difference between these systems
is that on a rough inclined plane,
the active layer flows down a substratum made of glued particles, while
in a rotating drum, the substratum, made of loose particles,
continuously exchanges particles with the flowing zone.
The incline with a gentle slope 
is an intermediate case between a rotating drum and a flow down a rough inclined plane
since a static bed can form in the basal part.
The incompatibility between these scaling laws leads to the development 
of theoretical models of granular flows that have to consider each case 
independently \cite{AradianRaphael00}.  
Another incompatibility appears if we consider the work of Parker et
al. \cite{ParkerDijkstra97} who report that in a drum, the thickness of the flowing layer
$h$ is constant with the rotation speed $\Omega$. Orpe and Khakhar also studied the 
flow of a dry granular material in a rotating drum \cite{OrpeKhakhar01}. 
In their study, they did
not directly extract a scaling law connecting velocity and thickness of the
flowing layer. But they measured
that the velocity gradient $\dot\gamma$ depends on the angle $\beta$ 
of the free surface ($\dot\gamma\sim \sqrt{\sin(\beta-\beta_s)}$, where $\beta_s$
is the angle of repose)
and that $\beta$ increases linearly with the rotation 
speed ($\beta-\beta_s\sim \Omega$). From these two relations,
and assuming that the angle difference remains small
($\sin(\beta-\beta_s)\simeq \beta-\beta_s$), one easily deduces 
the scaling law $v\sim h^3$. 
This result is intermediate between a constant velocity gradient, associated
to the scaling $v\sim h$
\cite{Rajchenbach00,BonamyDaviaud02} and a constant
thickness of the flowing layer $h\sim v^0$ \cite{ParkerDijkstra97} in a 
rotating drum. 

The work presented here reconsiders the problem of the scaling law
$v$ vs $h$ in a rotating drum. More precisely, the questions we address to
are: is there a scaling law connecting thickness and velocity of the flow in a 
rotating drum ? Is the exponent unique, or does it depends on the 
geometry ? For ``large drums", do we expect the same exponent as
for a rough incline plane whose curvature is infinite ? 
To answer these questions, a large
number of drums, sizes of beads and rotation speeds are studied. Indeed,
in previous work \cite{Rajchenbach00,BonamyDaviaud02,ParkerDijkstra97,OrpeKhakhar01}, 
the ratio diameter of the drum 
$D$ over diameter of the beads $d$ is always inferior to 400.
In this work, the ratio $D/d$ varies from 47 to 7400, while the rotation speed ranges
from 2 to 25 rpm. 
In our experiments, we measure the thickness of the flowing layer $h$
 versus the rotation speed $\Omega$. To obtain the link between the
power law $v\sim h^m$ and the law $h\sim \Omega^n$ that is 
deduced from our experiments, we use the definition of the flow rate $Q=h v$
where $v$ is the mean velocity of the flow, and note that the flux entering
the flowing layer in a half-filled rotating drum is given by $Q=\pi\Omega (R^2-h^2)$
 \cite{Rajchenbach00} where $R$ is the drum radius. Here
$h^2$ is assumed small compared to $R^2$ 
\cite{hsmall} and the previous relation is simplified in $Q=\pi\Omega R^2$. 
One easily obtains $m=(1-n)/n$. Thus, the scaling law previously reported, 
$v\sim h$, corresponds to $h\sim \Omega^n$ with $n=1/2$. 

\section{Experimental apparatus}
\begin{figure}[htbp]
\center
\includegraphics[width=\linewidth]{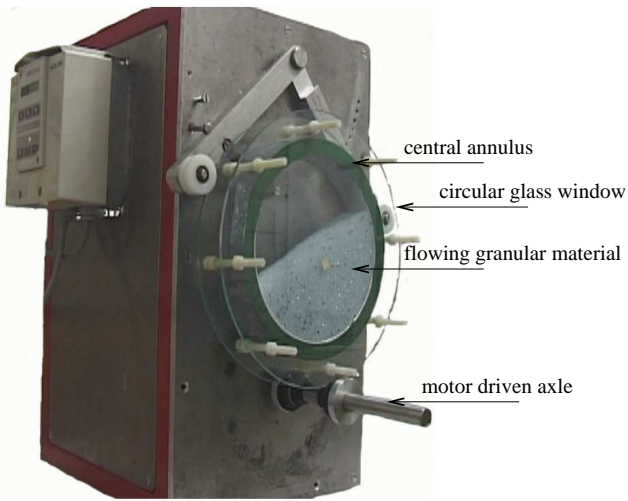}
\caption{Picture of the experimental apparatus}
\label{tambourschema}
\end{figure}
The drum is composed of an annulus of diameter $D$ (5.6, 12, 20, 30
and 50~cm) and width $W$ (0.6, 1, 2, 3, 4, and 6~cm), held between 
two circular glass windows, and placed vertically on a rotating axle 
driven by a continuous current motor (figure \ref{tambourschema}). The rotation speed of the drum $\Omega$ varies
from 2 to 25 rpm in order to be in the rolling regime (continuous flow regime).
The drum is half-filled
with glass beads of mean diameter $d$, ranging from 0.07~mm
to 2~mm, with a density of 2.5 g/cm$^3$. The studied size ratios (and mean 
beads diameter) are: 
$D/d$~=~47 ($d$~=~1.2~mm), 100 (560~$\mu$m, 1.2~mm, 2~mm), 166 (1.2~mm), 
214 (560~$\mu$m), 357 (560~$\mu$m), 600 (200~$\mu$m), 895 (335~$\mu$m), 
1000 (200~$\mu$m), 1500 (200~$\mu$m), 2500 (200~$\mu$m) and 7400 (75~$\mu$m). 
The case in which 45-90~$\mu$m glass beads are used
(ratio $D/d$~=~7400) has to be considered with care.
The corresponding results are reported since they are compatible with the others.
The humidity is held between 50\% and 55\% in order to reduce electrostatic
effects or capillary bridges between fine particles \cite{Fraysse99}. 
Experiments are filmed with a CCD camera whose 
shutter speed is chosen in order to follow
particle trajectories (long exposure time). This is analogous to the 
technique reported in Orpe and Khakhar \cite{OrpeKhakhar01}.
Thus the particles at the lowest point of the base of the flowing 
zone appear to be at rest on the film, defining a static point in the 
laboratory frame reference. $h$ is the distance measured between the static 
point and the free surface, perpendicularly to this surface.  
In fact, one would like to measure a static point in the drum frame
reference. But we will see that in the flow, the velocity decreases rapidly
to zero 
and reconnects with a plastic deformation zone. Thus there is no static point
in the drum frame reference. There are 2 possibilities to define a 
flowing layer: 
to assume a linear velocity profile to calculate a point where the velocity 
is equal to 0 in the drum frame reference \cite{GwenThese}, or to take the
static point in the laboratory frame reference.
The distance between these 2 points is very small 
and the choice does not affect the values of the results
 (thicknesses of the flowing layer, velocity profiles and exponents of
the scaling laws). 
In the following, we present the results obtained in the laboratory frame reference 
since they are directly measured in the experiment, with no assumed
hypothesis.

For each experiment, measurements
are reproduced between  10 and 20 times, which allows to represent mean data
with error bars 
using the Student test with a 95\% interval of confidence (standard deviation and
number of measurements are taken into account in the error bars).

The possible influence of several parameters has been considered: the 
precision of the filling of the drum (47.5, 50 and 53.5\%), the size distribution 
of the particles, the width of the drum ($D$~=~20~cm, $d$~=~2~mm, 
$W =$~0.5, 1, 2, 4, 6~cm) and the influence of the 
experimenter in the measurement of the static point (4 different 
experimenters). In all the cases, the discrepancy in the $h$ measurements is not 
statistically significant \cite{GwenThese}. Finally, we are aware of the fact that all measurements are made
through a lateral window, but several experimental works suggest that the 
thickness of the flowing zone at the wall is not, or only slightly,
modified compared to those at the center, even if the velocity is reduced
at the wall \cite{OrpeKhakhar01,GwenThese,ChevoirProchnow01,BoatengBarr97,TaberletRichard03} 

\begin{figure}[htbp]
\center
\includegraphics[width=\linewidth]{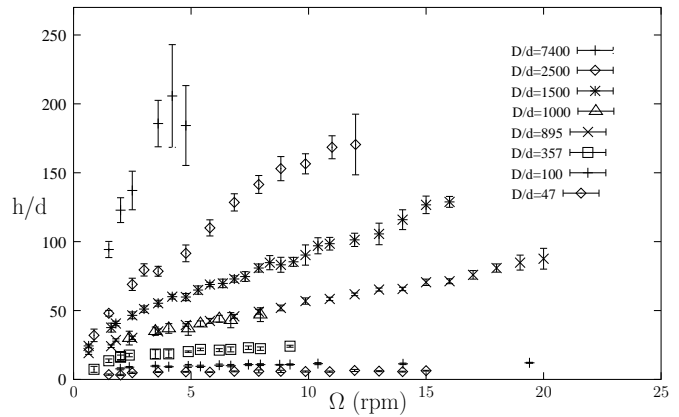}
\caption{Thickness of the flowing zone 
measured in bead diameters $h/d$ vs rotation speed $\Omega$ of the drum
for several size ratios $D/d$ (diameter of the drum $D$ over diameter of the 
beads $d$, the $D/d=100$ ratio corresponds to $d=2mm$)}
\label{fitallnew}
\end{figure}

\begin{figure}[htbp]
\center
\includegraphics[width=\linewidth]{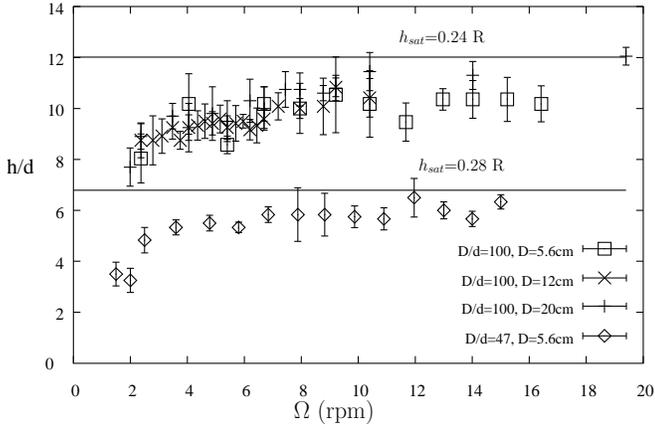}
\caption{Thickness of 
the flowing zone for $D/d$~=~47 and 100 
versus rotation speed $\Omega$. The 
two horizontal lines propose saturation thicknesses for the flowing zone 
arbitrarily chosen as the maximum data value.}
\label{fithomothetic}
\end{figure}

\section{Results: Thickness of the flowing zone}
Figure \ref{fitallnew} presents the thickness of the flowing zone (measured in bead 
diameters $h/d$) versus the rotation speed. The curves organise in raising 
ratios $D/d$.  For clarity reasons, all the studied ratios are not reported
on the graph, even if they all show the same behaviour \cite{GwenThese}.

One may note that for low size ratios (typically $100<D/d<400$), the thickness 
of the flow is close to a 10-beads layer. Even if this value is frequently 
reported, figure \ref{fitallnew} shows that this size of the flowing layer is not 
universal but depends on the
$D/d$ ratio. In our experiments, the thickness of the flow ranges from 4 
to 200 beads.

Special attention was put into homothetic systems, i.e. systems with  
equal size ratios, to check that $h/d$ and $D/d$ are the relevant parameters. 
For these homothetic systems ($D/d$~=~100, $W/d$~=~20 and
$D$~=~5.6, 12, 20~cm), all the curves superimpose within
the error bars (figure~\ref{fithomothetic}). This has already been reported in 
\cite{GwenThese,GwenPowdGrains} while studying the angle of avalanches, of 
repose and of continuous flow. 
This can also be observed on figure~\ref{fitallnew} for close ratios ($D/d$~=~895, 
$D$~=~30~cm and $D/d$~=~1000, $D=$~20~cm). This result suggests that the Froude
number, Fr=$\omega^2 R/g$ with $\omega=2\pi\Omega$ and $g$ the gravity 
acceleration, is not the relevant parameter to describe the
thickness of the flowing zone. Indeed, in the cases $D/d$~=~895  and
$D/d$~=~1000, if the thickness is plotted versus Froude number, the curves do not
superimpose as precisely as in figure~\ref{fitallnew}.

The flowing layer thickness curves presented on figure~\ref{fitallnew} show two
different behaviors:
for low ratios $D/d$, the 
flowing zone increases and rapidly reaches a constant flowing thickness (saturation thickness).
Parker et al. \cite{ParkerDijkstra97} report a saturation thickness corresponding 
in average to 35\% of the radius $R$ of the drum (but with large variations), with size ratios ranging
from $D/d$~=~33 to 90. Figure \ref{fithomothetic} shows that in our experiments, the saturation thickness is around
28\% (resp. 24\%) of the radius of the drum $R$ for a size ratio $D/d$~=~47 
(resp. $D/d$~=~100).  
For higher ratios, the flowing layer
thickness increases continuously with the rotation speed $\Omega$. 
For these size ratios, one expects that $h$ would stop increasing when a
saturation thickness will be reached. 
But we assume this would occur for
rotation speeds high enough such that other phenomena might have already 
appeared: strong ``S" shaped free 
surface, cataracting and eventually centrifugation.

\section{Saturation regime}

In the saturated regime, the
thickness of the flow remains almost constant when $\Omega$ increases, as previously observed 
\cite{ParkerDijkstra97}. The
increase of the flow rate is mainly adapted by an increase of the mean velocity. 
We deduced that, in this regime, the velocity gradient cannot be constant, and should increase
with the rotation speed.

As this deduction is indirect and concerns a gradient averaged on the whole 
flowing layer, 4 velocity profiles $v_x(y)$ have been measured along an axis 
($Y$) perpendicular to the free surface (defining the $X$ axis). This
will allow us to compare velocity profiles and their associated scaling 
law with the one obtained by previous authors \cite{Rajchenbach00,BonamyDaviaud02}.
A high speed camera (800 fps) have been used.
The shift of
one particle between two frames is about a few bead diameters.
Thus, we reject the
idea of using a particle image velocimetry software. Instead, we follow the
trajectories of a few colored glass beads, 
selecting them ``by hand" on a specially dedicated software. 

\begin{figure}[htb]
\center
\includegraphics[width=\linewidth]{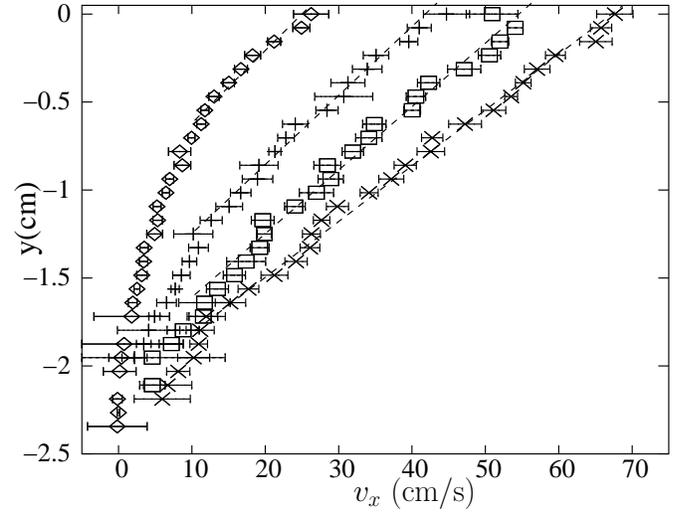}
\caption{Velocity profiles measured in a rotating drum $D$ =~20~cm 
with a size ratio $D/d=100$, $y$ is the position along an axis perpendicular to the free surface
$y$~=~0, and fits on the linear part of each profile.
The error bars are obtained using a Student test with a 95\% interval.
 The experiments are performed with 4 rotation speeds 
(from left to right $\Omega=$~4, 9.2, 14, 19.4 rpm). The 
velocity gradient in the linear part increases with the rotation speed.}
\label{velocityprofileexp}
\end{figure}

\begin{figure}[htb]
\center
\includegraphics[width=\linewidth]{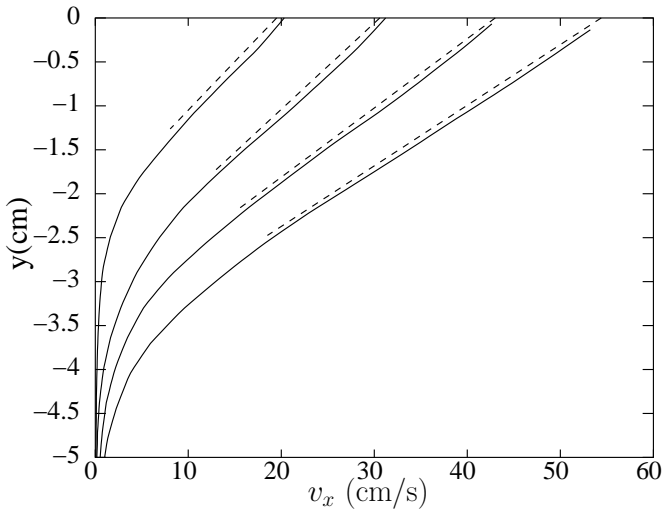}
\caption{Velocity profiles measured in a rotating drum $D$ =~20~cm 
with a size ratio $D/d=100$, $y$ is the position along an axis perpendicular to the free surface
$y$~=~0, and fits on the linear part of each profile.
The numerical simulations are in 2D and performed with 4 rotation speeds (from left to right 
$\Omega=$~5, 10, 15, 20 rpm).
Like in experiments, the 
velocity gradient in the linear part increases with the rotation speed.}
\label{velocityprofilesimu}
\end{figure}

Figure \ref{velocityprofileexp} shows velocity profiles measured
for rotation speeds of $\Omega=$~4, 9.2, 14, 19.4~rpm ($D/d=$~100, $D=$~20~cm).
Between 630 and 1200 velocities have been measured for each of the four profiles.
As pointed out by Komatsu et al. \cite{KomatsuInagasaki01}, the profiles
display a linear part and an exponential part 
(reconnection with the static zone). 
For the 3 higher rotation speeds (figure \ref{velocityprofileexp}), we 
can clearly see a linear part in the velocity profiles, while for the 4~rpm 
case, the overall profile seems convex.


The increase of the total velocity gradient 
can be induced by two effects: an increase of the slope of the linear part, and 
also an increase of the extension of this linear part, and thus a decrease
of the reconnection zone thickness.

From figure \ref{velocityprofileexp}, we see that the linear part increases strongly between
$\Omega=4$~rpm and $\Omega=19.4$~rpm and this contributes to an increase of the 
total velocity gradient of the flowing layer.
We will now focus on the linear part, and see if it presents a constant velocity
gradient as previously reported \cite{Rajchenbach00,BonamyDaviaud02}.  Although the linear part is not obvious for
the 4~rpm case, velocity gradients are obtained by fitting the upper part 
(dashed lines) of the velocity profiles. 
The obtained velocity
gradients are not equal ($\dot \gamma$~=~24.2, 25.5, 27.7, 31.5~s$^{-1}$),
 but increase with the rotation speed. 

As a fit of the linear part might be subject to discussion (especially the frontier between
the linear part and the exponential reconnection is rather subjective), we 
have decided to perform 
numerical simulations with a geometry close to the experimental conditions: 4300 disks
of 2~mm diameter flowing in a 2D drum of 20 cm diameter. The numerical method 
is a distinct element method, with a linear spring dash-pot model for the normal 
forces and the Cundal and Strack scheme for the tangential forces (see for
example \cite{DippelWolf96}). More details on the numerical simulations
will be given elsewhere \cite{DOrtona05}. Four rotation speeds are simulated 
$\Omega$~=~5, 10, 15 and 20 rpm. The linear part and the 
exponential reconnection with the static part appear more clearly (figure \ref{velocityprofilesimu}), and 
the measured velocity gradient does not depend on the frontier with the
reconnection zone (while remaining in the dashed lines region).
Again, the extension of the linear part increases with the rotation speed $\Omega$.
Moreover, the velocity gradients obtained by fitting the linear part
increase with the rotation speed (resp. $\dot \gamma=$~8.23, 10.57, 12.61  
and 14.63~s$^{-1}$). The values
obtained in the simulation are different from the experiment,
probably due to the fact that the simulations are in two dimensions.

We see that, in both cases (experiment or simulation), 
the velocity gradients taken on the whole layer, including
the reconnection zone (global), or taken on the linear part of the profile 
(local) are not constant and increase with the rotation speed. Thus,
if a scaling law $v\sim h^m$ is used to fit our data on velocity
and flowing thickness, the exponent $m$ would be superior to~1. 


\section{Velocity scaling laws}
From our experiments, we deduce scaling laws
between the thickness of the flowing zone $h$ and
the rotation speed $\Omega$. We then derive scaling laws
connecting thickness of the flowing layer $h$ and mean velocity $v$ and 
compare with the results obtained by previous authors. 
\begin{figure}[htb]
\center
\includegraphics[width=\linewidth]{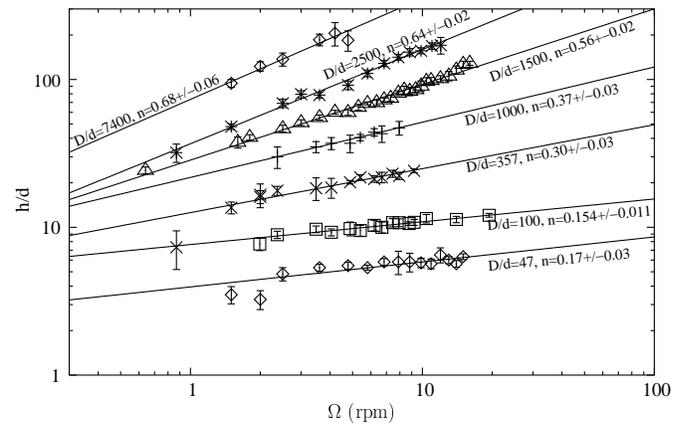}
\caption{Thickness of the flowing zone of the granular material 
measured in bead diameters $h/d$ vs rotation speed $\Omega$ of the drum
for several size ratios $D/d$; fits with a power law $h\sim \Omega^n$ and values of 
the exponent $n$ with their uncertainty. The exponent $n$ is
 not constant and increases with the size ratio.}
\label{fitlog}
\end{figure}

Figure \ref{fitlog} shows the experimental data for several values of $D/d$ in a
log-log plot
and crude fits using a power law $h\sim \Omega^n$ with a
least square method.
When a power law is used to fit these data, very good
regressions are obtained. We found that the obtained exponent is not unique, 
and depends on the $D/d$ ratio.  
Table 1 gives the exponents obtained by fitting $h$ vs $\Omega$
and the corresponding power law $v \sim h^m$. For the small ratios $D/d$, the $m$
 exponents are much larger than 1, which is in accordance with the increase
of the global velocity gradient with $\Omega$ (fig. \ref{velocityprofileexp}).
For example, the power law for the ratio 
$D/d$~=~100 is $h\sim \Omega^{0.154\pm 0.011}$, and induces a ``pseudo'' scaling 
$v\sim h^{5.5}$ never reported before. 
When the 
size ratio $D/d$ increases, this exponent $m$ continuously decreases, with
values inferior to 1 for the 3 cases $D/d$~=~1500, 2500 and 7400.

\begin{figure}[htb]
\center
\includegraphics[width=\linewidth]{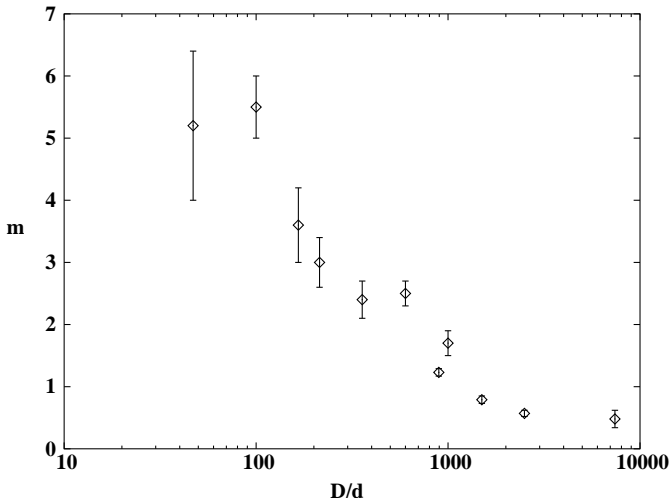}
\caption{Evolution of the exponent $m$ of the power law $v\sim h^m$ with
the size ration $D/d$. The exponent decreases continuously with the size ratio.}
\label{exponent}
\end{figure}

%
For small size ratios ($D/d< 100$), to fit the data using a power law
might seem not logical since we just shown that
a saturation thickness is expected for high rotation speeds. A function with 
an
asymptotic behavior should be a better choice. Our aim here is simply to
show that the existence of a saturation thickness does not appear through 
a sharp
transition in the curves: the saturation process
happens progressively, affecting the 
thickness of the flow even for low rotation speeds. For larger size ratios,
this saturation process probably also influences the thickness of the flow, even
if the saturation thickness is not reached. Indeed, figure \ref{fitlog} shows that 
the slopes $h/d\sim \Omega$ continuously increase. 
These fits show that if a universal power law between
$h$ and $\Omega$ (or $h$ and $v$) exists in a rotating drum, it can not be obtained
for small size ratios, but only in systems with a flowing 
zone thickness small compared to the saturation thickness. One solution 
might be to use low rotation speeds to get thin flowing layers.
Unfortunately, the system reaches the avalanching regime, and the flow is no 
longer continuous. The other solution is to study high size ratios, expecting
that the exponent $m$ would reach an asymptotic value. But in our experiments,
no asymptotic value is obtained. When increasing the size ratio, the exponent 
$m$ decreases continuously, taking values
close to 1 for size ratio $D/d$ between 1000 and 1500, and values lower
than 1 for our larger size ratios (see figure \ref{exponent} and table \ref{tablexposant}).
\begin{table}
\center
%
\begin{tabular}{|l|l|l|}
\hline
$D/d$ & $h\sim \Omega^n$ & $v\sim h^m$ \\
\hline
47 & 0.17$\pm$0.03 & 5.2$\pm$1.2 \\
 100& 0.15$\pm$0.011& 5.5$\pm$0.5 \\
 166& 0.22$\pm$0.03& 3.6$\pm$0.6 \\
 214& 0.25$\pm$0.03& 3.0$\pm$0.4 \\
 357 & 0.30$\pm$0.03 &2.4$\pm$0.3 \\
 600 & 0.28$\pm$0.02 &2.5$\pm$0.2 \\
 895 & 0.449$\pm$0.013& 1.23$\pm$0.06\\
 1000 & 0.37$\pm$0.03& 1.7$\pm$0.2\\
 1500 & 0.56$\pm$0.02& 0.79$\pm$0.06\\
 2500 & 0.64$\pm$0.02& 0.57$\pm$0.05\\
 7400 & 0.68$\pm$0.06& 0.48$\pm$0.14\\
\hline
\end{tabular}
\caption{Exponents of the power law $h$ vs $\Omega$ obtained in the 
experiments and their equivalent exponents of the power law  
$v$ vs $h$ obtained using $m=(1-n)/n$}
\label{tablexposant}
\end{table}
\begin{figure}[htb]
\center
\includegraphics[width=\linewidth]{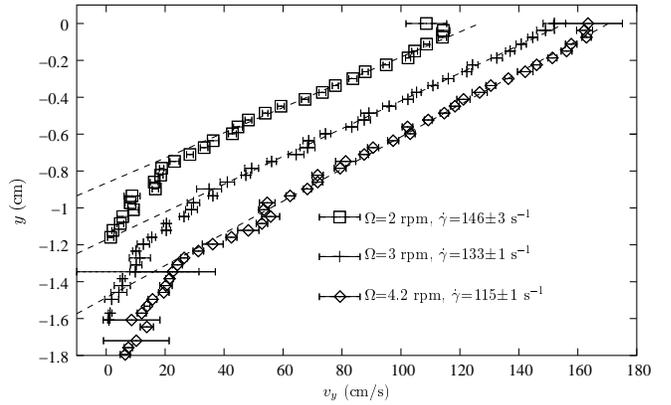}
\caption{Velocity profiles measured in a rotating drum $D$~=~50~cm with 
a size ratio $D/d=1500$ for 3 rotation speeds ($\Omega$~=~2, 3, 4.2 rpm), fits
on the linear part of the profiles, and deduced velocity gradients $\dot \gamma$ .
The velocity gradient decreases with an increase of the rotation speed.}
\label{velocityprofilelarge}
\end{figure}

An exponent $m$ lower than 1 induces a decrease of  the global velocity 
gradient when 
rotation speed increases. 
This result is opposite to what is observed
for small size ratios. To confirm this counter-intuitive 
evolution, direct velocity profiles measurements were performed.
The system $D/d$~=~1500 was chosen as a compromise between
the feasibility of measurement (the particles are still distinguishable) and a
power law with an exponent smaller than 1. Figure \ref{velocityprofilelarge} shows velocity profiles
with the associated error bars. 
The error bars are smaller than on
figure \ref{velocityprofileexp} because between 2500 and 4100 particle displacements have been 
measured for each profile. This was necessary 
to reduce the error in the velocity gradient estimation, and to have
statistically significant differences in the slope of the profiles.
Again the linear part growths. But, this time, the evolution of the 
reconnection zone is not clear. Thus we cannot 
explain the obtained power law only by these two evolutions. 
On the other hand, 
the velocity gradients in the linear part have
been measured slightly decreasing with $\Omega$ which is compatible with
a power law inferior to 1. 
In both cases (figure \ref{fitlog} and \ref{velocityprofilelarge}), the global velocity gradients
deduced from the power law and the local gradients
measured on the linear part of the profiles evolve jointly. 

%

\section{Discussion}
These experiments have shown that both global and local velocity gradients of 
a granular flow in a 
rotating drum are not constant with the rotation speed. The variation of 
the local velocity gradient is new compared to
results reported before
\cite{Rajchenbach00,BonamyDaviaud02},
even if an increase of the velocity gradient for high rotation speed
is also suggested by the data (Fig. 20) of Orpe and Khakhar 
\cite{OrpeKhakhar01}. Our results induce that the
scaling law $v\sim h$ is not universal in a drum but that the value of $m$ in $v\sim h^m$
 depends on the size ratio between drum and beads.
In our experiments, the peculiar power law $m=1$ should be obtained for
a size ratio $D/d$ comprised between 1000 and 1500. 
In their experiments, Rajchenbach \cite{Rajchenbach00} and Bonamy et al. 
\cite{BonamyDaviaud02} have obtained this
scaling for size ratios $D/d$ around 150. The difference is 
probably due to the fact that Rajchenbach reports the use of steel
beads in a 2D system, while Bonamy et al. have used steel and glass beads, but
in a quasi 2D system. We also note that our scaling laws are obtained with
the measurement of the thickness of the flow including a part of the
reconnection zone while Rajchenbach and Bonamy et al. found that the 
velocity gradient of the linear part is constant.

Recently, Taberlet et. al. \cite{TaberletRichard03} show that the thickness of
the flowing layer of a heap in a thin channel is imposed by the channel width.
Jop et. al. \cite{JopForterre05} also show that side walls control the steady flow on pile for
channel width  ranging from 20 to 600 particle diameters.
But in our experiments, the width can not determine the thickness of the flowing layer
since it was shown that changing the width of the drum
 does not affect the thickness of the flowing zone
\cite{GwenThese}. In our system, 
the thickness of the flowing layer is mainly imposed by the size ratio (diameter
of the drum on diameter of the beads)
and by the rotation speed (probably due to the saturation process). 




\section{Conclusions}

The flow of dry granular material (diameter $d$) in a rotating drum (diameter
$D$) with size ratios $D/d$ ranging from 47 to 7400 is studied. The thickness
of the flowing layer $h$ is measured for increasing rotation speeds $\Omega$.
Experimental curves corresponding to the same ratio $D/d$ coincide 
suggesting that this ratio is the leading parameter (with the rotation speed)
imposing the thickness 
of the flowing zone in a rotating drum. 

For small size ratios, a saturation thickness, previously reported in literature,
is observed, inducing a thickness of the flowing zone constant with the 
rotation speed, and thus a velocity gradient increasing with the rotation speed.

If a power law $v\sim h^m$ is deduced from these experiments, the exponent
$m$ is not constant as previously reported, but varies from 5 
(small size ratios) to 0.5 (larger size ratios).  An exponent
smaller (resp. larger) than 1 induces a decrease (resp. increase)  of the 
velocity gradient while the rotation 
speed increases. This has been confirmed by direct measurements of the velocity
profile.

We also see that for large
size ratios, the exponent does not tend to the value obtained for a flow
down a rough incline plane: $v\sim h^{3/2}$. This shows that the two 
systems, flow down an incline and rotating drum, are fundamentally different. 
The difference is probably due to the presence or not of a static bed of beads.
The question that remains to be answered is: when increasing the size ratio
($D/d$), does the exponent of the scaling law $v\sim h^m$ tends
to 0 like observed by Ancey \cite{AnceyPRE02} for a flow on an incline
with a basal static bed, or to some non null value, like $m=1/2$, as suggested
by data in our larger system $D/d$~=~7400 ?

\begin{acknowledgement}
We wish to acknowledge G\'erard Verdier for his image processing software and
 Nathalie Thomas for interesting discussions and careful reading of this article.
\end{acknowledgement}


\begin{thebibliography}{99}
\bibitem{hsmall} Taking into account the thickness of flowing
layer $h$ while deducing the exponent $n$ requires the fit of the 
mean velocity data that are not directly measured, and thus present a larger 
dispersion \cite{GwenThese}.  The exponents obtained by this method are 
close to those presented in table \ref{tablexposant}.
\end{thebibliography}
\end{document}